\documentclass[%
superscriptaddress,
preprint,
 bibnotes,
 amsmath,amssymb,
 aps,
]{revtex4-1}

\usepackage{graphicx}
\usepackage{dcolumn}
\usepackage{bm}
\usepackage[usenames,dvipsnames]{xcolor} 
\usepackage[colorlinks,citecolor=MidnightBlue,linkcolor=Orange,urlcolor=MidnightBlue]{hyperref}
\usepackage[mathlines]{lineno}
\usepackage{color}
\usepackage{MnSymbol}
\definecolor{orange}{rgb}{1.0,0.4,0.0}
\definecolor{rose}{rgb}{1.0, 0.33, 0.64}
\usepackage{tikz}
\usepackage{times} 
\usepackage{rotating}
\usepackage{gensymb}
\usepackage{siunitx}
\usepackage{xr}
\usepackage{commath}
 \usepackage{hyperref}
 \usepackage{cleveref}

\begin{document}

\newcommand{\blueline}{\raisebox{2pt}{\tikz{\draw[-,black!10!blue,solid,line width = 1pt](0,0) -- (3mm,0);}}}

\newcommand{\blackline}{\raisebox{2pt}{\tikz{\draw[-,black!40!black,solid,line width = 1pt](0,0) -- (3mm,0);}}}

\newcommand{\magentaline}{\raisebox{2pt}{\tikz{\draw[-,black!5!rose,solid,line width = 1pt](0,0) -- (3mm,0);}}}

\title{Segregation and cooperation in active colloidal binary mixtures}

\author{Laura Alvarez}
\email{laura.alvarez-frances@u-bordeaux.fr}
\affiliation{Univ. Bordeaux, CNRS, CRPP, UMR 5031, F-33600 Pessac, France}%
\affiliation{Laboratory for Soft Materials and Interfaces, Department of Materials, ETH Zurich, 8093 Zurich, Switzerland}%
\author{Elena Ses\'e-Sansa}%
\affiliation{American Physical Society, 100 Mtr Pkwy, Hauppauge, New York 11788 (USA)}
\author{Demian Levis}
\email{levis@ub.edu}
\affiliation{Computing and Understanding Collective Action (CUCA) Lab, Condensed Matter Physics Department, Facultat de Física, Universitat de Barcelona, Mart\'i i Franqu\`es 1, E08028 Barcelona, Spain}%
\affiliation{University of Barcelona Institute of Complex Systems (UBICS), Mart\'i i Franqu\`es 1, E08028 Barcelona, Spain}
\author{Ignacio Pagonabarraga}
\affiliation{University of Barcelona Institute of Complex Systems (UBICS), Mart\'i i Franqu\`es 1, E08028 Barcelona, Spain}
\affiliation{Condensed Matter Physics Department, Facultat de Física, Universitat de Barcelona, Mart\'i i Franqu\`es 1, E08028 Barcelona, Spain}%
\author{Lucio Isa}%
\affiliation{Laboratory for Soft Materials and Interfaces, Department of Materials, ETH Zurich, 8093 Zurich, Switzerland}%

\begin{abstract}

The complex interactions underlying collective motion in biological systems give rise to emergent behaviours such as flocking, sorting, and cooperative transport. These dynamics often involve species with different motilities coordinating movement to optimize navigation and survival. Synthetic analogues based on active colloids offer a controlled platform to explore such behaviours, yet most experimental realizations remain limited to monodisperse systems or mixtures of passive and active particles. Here, we investigate dense binary mixtures of active Janus colloids with distinct motilities and independently tunable alignment, actuated by AC electric fields. We demonstrate experimentally and numerically that both species form highly dynamic polar clusters, with alignment emerging independently of propulsion speed. In mixed populations, interspecies interactions lead to effective segregation and cooperative motion, including transient enhancement of slower particle motility. Our results reveal how motility contrast and alignment combine to drive self-organization in active mixtures, offering strategies for designing reconfigurable materials with collective functionalities.

\end{abstract}
\maketitle

\section*{Introduction}

Group motility is ubiquitous in biological systems across various length scales. This dynamical feature arises due to the presence of intricate interactions and of external stimuli, as observed at the microscale in densely populated bacterial \cite{Zhang2010,Rombouts2023,Burriel2024} or amoeba \cite{Gregor2010} colonies, actin filaments \cite{Schaller2010,Suzuki2017} and skin cells \cite{Puliafito2012}. Moreover, populations of active biological units with different motilities may exhibit segregation \cite{Grobas2021,Anderson2022}, causing spontaneous spatial sorting amongst species, either via direct competitive mechanisms \cite{Zuo2020,Rosenberg2016}, or by altruistic cooperation of different species to optimize survival \cite{Ebrahimi2019}.  

Some of the complex collective dynamics of biological entities is nicely captured by self-propelling, or active, colloidal particles, which constitute a model system to single out the essential underlying physics of dense out-of-equilibrium collectives  \cite{Ebbens2016,Zottl2016,Dauchot2022}. These artificial microswimmers convert external sources of energy into net propulsion leading to emergent collective behaviour at high particle population density. A plethora of phenomena arise by tuning the particle-particle interactions and motilities of the active units, usually done by carefully choosing the actuation mechanism and materials forming the active particles \cite{Zottl2016}. Self-phoretic active colloids, for instance, self-assemble into dense clusters with low motilities, within a gas phase of freely-moving particles \cite{Theurkauff2012,buttinoni2013dynamical, Palacci2013, Singh2017}. From a theoretical standpoint, such a phenomenon is understood in the context of the so-called motility-induced phase separation \cite{Cates2015}: a condensation transition arising from the mere competition between self-propulsion and excluded volume interactions \cite{tailleur2008statistical, redner2013structure, speck2015dynamical, digregorio2018full}, without any collective alignment or polar order.
Conversely, Janus colloids self-propelling via AC electric fields experience aligning events via hydrodynamics and dipolar interactions, depending on how the field is applied and the particle asymmetry. This actuation mechanism, therefore, leads to the emergence of rich group behaviour such as the formation of particle chains, chiral clusters \cite{Zhang2020,Zhang2021,vanBaalen2025}, flocks with high global orientational order \cite{Bricard2013,Iwasawa2021,Katuri2022}, or very dynamic small polar clusters with short-range orientational order \cite{Karani20219,Nakayama2023}. Vicsek-like models capture this particle alignment in a minimal way \cite{Vicsek1995, chate2020dry} assuming velocity-alignment interactions between such self-propelled particles. 

While the collective behaviour of ensembles of particles with a single motility has been extensively explored, the emerging group dynamics of mixtures of particles with different motilities remains mostly untapped. Experimentally, mixing active particles into passive particle systems has been shown to affect the dynamics and structure of the latter \cite{Kummel2015,Gokhale2022,Mauleon-Amieva2020,MassanaCid2018,Kushwaha2023,Burriel2023,Zion2022}. In particular, as theoretical work suggests, it may increase their diffusivity, i.e. acting as a higher effective temperature, or lead to trapping of passive particles in dense clusters of active ones \cite{Stenhammar2015,Wittkowski2017,Wysocki2016,vanderMeer2020,RogelRodriguez2020}. However, experiments where two active species of the same size but different motility are mixed and self-propel under the same conditions, have been difficult to realize.
This case has been recently explored in numerical simulations of non-aligning self-propelling particles, with a particular focus on the motility-induced phase separation regime \cite{Kolb2020}, upon varying the activity and concentration of each species. This latter work extends the current kinetic theory of active/passive mixtures to active/active mixtures, drawing an interesting analogy to equilibrium systems of passive particles with two effective temperatures \cite{Grosberg2015,Weber2016}. 
Despite recent advances, experimental studies of mixed populations of self-propelling and aligning particles remain scarce. Understanding the collective behavior of active particle mixtures could pave the way for designing active materials with dynamic reconfigurability, or targeted transport capabilities. These insights could help in the design of complex active materials with selective self-organization and 
via cooperative or competitive behavior, towards bioinspired "social" dynamics, where non-reciprocity plays an important role \cite{Gompper2020,Boudet2021,dinelli2023non, Hallatschek2023}.

In this work, we study the emergent collective behaviour of mixed populations of Janus colloids of the same size but with two distinct motilities - fast and slow - at varying densities in the fluid regime under AC electric field actuation. We experimentally and numerically elucidate the rich dynamical behaviour deriving from polar interactions, which can be tuned thanks to the choice materials and experimental conditions, showing clustering, collective motion and species segregation. 

\section*{Results}

\subsection{Janus colloids with frequency-dependent polar alignment }

We design two types of Janus spheres with the same size but different swimming velocities $\rm v$ and similar reorientation times $\rm \tau_R$ under the same AC electric field conditions, i.e. at a fixed applied frequency $\rm f$ and peak-to-peak voltage $\rm V_{pp}$. 
Briefly, we sputter-coat a monolayer of silica ($\rm SiO_2$) spheres with a diameter of $\sigma = \rm 2~\mu m$ with (i) 2 nm of chromium (Cr) and 6.5 nm of palladium (Pd), and (ii) 2nm of Cr and 4 nm of gold (Au), respectively. The difference in the layer thickness between Au and Pd accounts for the difference in the metal densities $\rm \rho$ ($\rm \rho_{Au} > \rho_{Pd}$), to avoid different cap weights for the two-particle species (Supplementary Fig.S1). 
The resulting Janus particles are dispersed in MilliQ water and confined between two transparent electrodes consisting of two glass slides coated by a Cr and a Au layer, covered with a thin layer of $\rm SiO_2$ to reduce particle sticking, and separated by a 120 $\rm \mu m$ spacer (see Fig.\ref{fig:fig1}a, and Methods). 
The electrodes are connected to a function generator that applies a perpendicular AC electric field between 1 and 10 $\rm V_{pp}$ (field strength 8333-83333 Vcm$^{-1}$),  and frequencies between 0.8 to 3 kHz. 
In the kHz regime, both metal-coated Janus particles self-propel with the $\rm SiO_2$ hemisphere in front (see Fig.\ref{fig:fig1}a, and Supplementary Movie S1) \cite{Shields2017a,Ma2015}. 
The self-propulsion and interactions of Janus particles actuated by AC electric fields in the kHz regime, close to a bottom electrode, is driven by induced charge electro-osmotic (ICEO) flows \cite{Ganwal2008,Peng2014}. In this scenario, the unequal polarization of the metal-coated and dielectric hemispheres induces unbalanced fluid flows that lead to particle self-propulsion \cite{Shields2017a}.

\begin{figure}
\label{fig:fig1}
\includegraphics[width=1\linewidth]{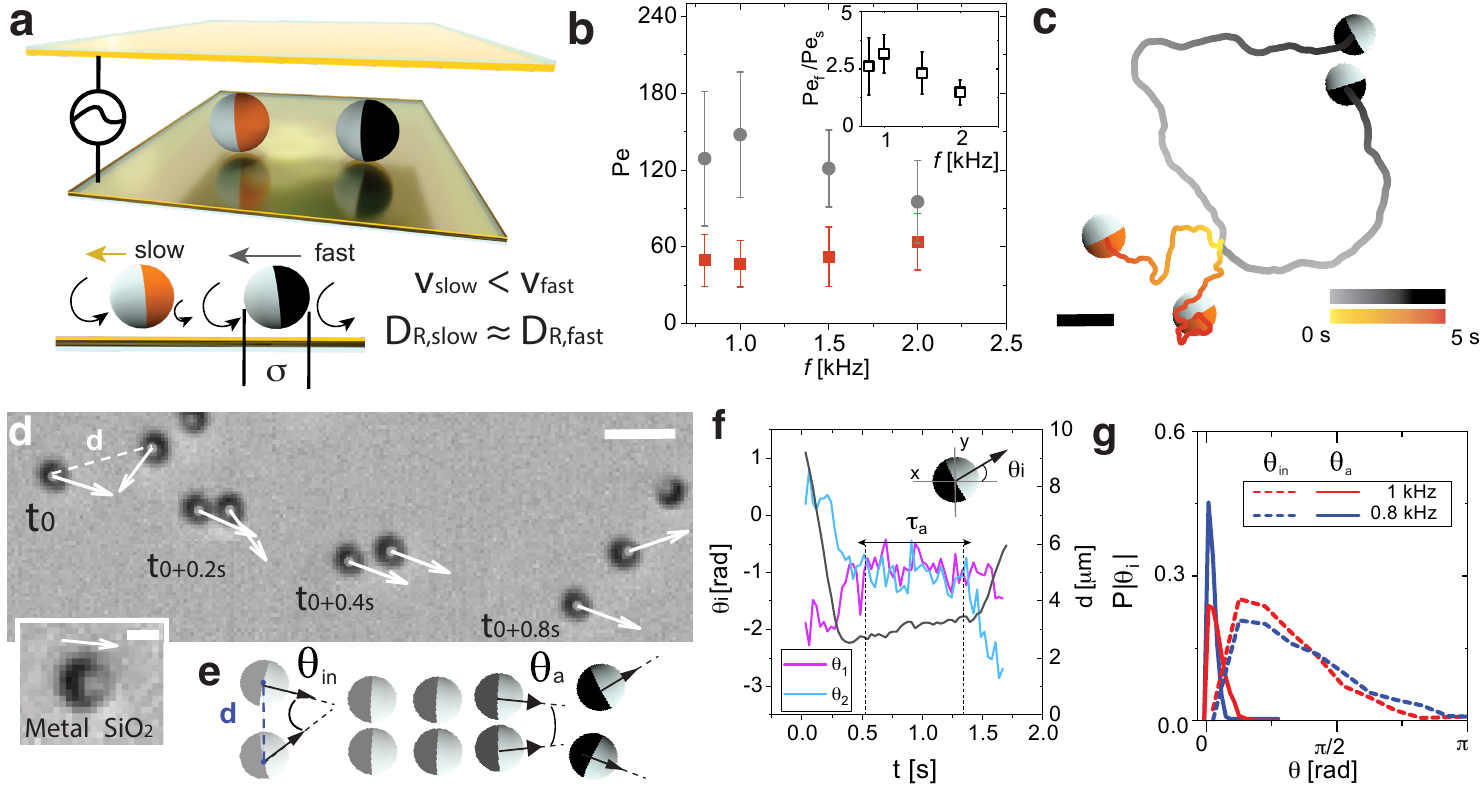}
\caption{\textbf{Active Janus spheres with two swimming speeds and aligning interactions.}~\textbf{a}, Schematic representation of the experimental cell, illustrating the application of the transverse AC electric field $E$, and two different Janus $\rm SiO_2$ spheres, i.e. Au-coated ($\rm SiO_2$-Au, orange) and Pd-coated ($\rm SiO_2$-Pd, gray) with diameter $\sigma$. Below, the electrohydrodynamic flows (black arrows) formed around each hemisphere of each type of particle are schematically shown. The arrows depict the velocity vector of each Janus colloid.~\textbf{b}, Péclet number $\rm Pe = \frac{v}{\sigma D_R}$, as a function of the frequency of the applied field applied for $\rm SiO_2$-Au (orange) and $\rm SiO_2$-Pd  (grey). The inset represents the ratio of Pe between the fast (Pe$_f$) and the slow (Pe$_s$) particles.~\textbf{c}, Representative trajectories for two $\rm SiO_2$-Au (orange) and two $\rm SiO_2$-Pd (gray) particles. The colour coding indicates time and the scale bar corresponds to 20 $\rm \mu m$.~\textbf{d}, Snapshots of an alignment event between two Janus particles at 1kHz and 8 $\rm V_{pp}$. The arrows indicate the orientation of the velocity vector, and the white dashed line is the inter-particle distance $d$. The scale bar indicates 5 $\rm \mu m$. The inset is a zoom-in to a Janus particle showing the $\rm SiO_2 $ and the metal hemisphere (light and dark, respectively). The scale bar corresponds to 1 $\rm \mu m$.~\textbf{e}, Schematic of two Janus particles approaching with an incoming angle $\theta_{in}$. Upon interacting, they acquire an alignment angle $\rm \theta_a$.~\textbf{f}, Relative Janus particle orientation $\rm \theta$ and interparticle distance $d$ as a function of time for a pair of particles undergoing an alignment event at 1 kHz and 8 $\rm V_{pp}$. The alignment time is defined as $\tau_a$. The inset depicts the relative orientation $\rm \theta_i$ with respect to the axis of reference x. ~\textbf{g}, Probability of particle orientations at 1 and 0.8 kHz (red and blue respectively) before (P($\rm \theta_{in}$, dashed line) and after (P($\rm \theta_{a}$, solid line) an alignment event.}
\label{fig:fig1}
\end{figure}

For each type of Janus particle, we analyze the single-particle dynamics as a function of frequency at a fixed applied voltage amplitude of 8 $\rm V_{pp}$, starting from dilute conditions at area fraction $A \approx$ 1\% (Supplementary Movie S1). We measure the self-propulsion speed $\rm v$ from the distribution of displacements over two-frame steps for each particle (dt$\rm  =  $0.06 s), and the rotational diffusivity $\rm D_R = {\tau_R^{-1}}$ from the relaxation of mean squared displacements (MSD) from ballistic to diffusive \cite{Bailey2022}. We use more than 1000 particles for each particle type, with trajectories having between 400 and 1000 data points each. 
The analysis indicates that $\rm SiO_2$-Pd particles are faster than $\rm SiO_2$-Au ones under the same conditions. As expected from theory \cite{Ristenpart2007, Ma2015}, the self-propulsion speed for each particle increases quadratically with the applied field, $\rm v\propto E^2$. We confirm that D$\rm _R$ does not depend on the applied field and is the same for both particles, i.e. D$\rm _R=$~0.28 $\pm$ 0.04\,$\rm rad^2s^{-1}$ (Supplementary Fig.S2). 
Once these quantities are known, we determine the Péclet number $\rm Pe=\frac{v}{\sigma D_R}$ (being $\rm \sigma$ the colloid diameter), as a function of voltage and frequency (Fig.\ref{fig:fig1}b,c) for each particle type. We find that the particles exhibit the highest difference in $\rm Pe$ at a voltage of 8 $\rm V_{pp}$ and a frequency of 1 kHz, with $\rm Pe^{exp}_{fast}$ ($\rm v_{F} =$ 62 $\pm$ 17 $\rm~\mu m s^{-1}$) approximately 2.5 times bigger than $\rm Pe^{exp}_{slow}$ ($\rm v_{S} = $24 $\pm$ 5~$\rm \mu m s^{-1}$).  These conditions will be used later in this study. While the $\rm SiO_2$ hemisphere is the same in both particles, the difference in swimming velocity arises from the difference in surface conductivity ($\sigma_S$) between Pd and Au, which leads to the difference in Pe.

While varying the frequency in our experimental range has a moderate effect on the swimming velocity for both particle types, we report an important effect on the alignment interactions between particles (Fig.\ref{fig:fig1}d,e). Thus, we study the occurrence of alignment events at low area fractions (A\% $<$ 0.1) at frequencies from 0.8 to 2 kHz. We define an alignment event using the particle's centre-to-centre distances $d$  when two particles come closer than $d \approx 1.2 \times \sigma$, during an alignment time  $\rm \tau_a>$ 0.1 s (see Fig.\ref{fig:fig1}f and Supplementary Movie S2). This threshold is based on measurements of $\rm tau_a$ and $d$ during particle collisions in the disordered phase as a reference, where there is no alignment. Moreover, we measure the distributions of incoming angles $\rm \theta_{in}$ before collision, and angles during the alignment $\rm \theta_{a}$ (Fig.\ref{fig:fig1}g). At 2 kHz, we do not observe any alignment as particles interact, and they only experience gas-like collisions (Supplementary Movie S2). For frequencies below 2 kHz, we observe alignment events with increasing $\rm \tau_{a}$ and the interparticle distance of a pair of aligned particles decreases $d$ (Supplementary Fig.S3). The results indicate a rather broad range of orientations for particles entering a collision $\rm \theta_{in}$ before alignment; while the distribution of angles during alignment events $\rm \theta_{a}$ is narrower and closer to zero at lower frequency, indicating a stronger alignment during contact.

Previous studies have attributed alignment and particle attraction to dipole-dipole interactions in the MHz regime \cite{Yan2016, Iwasawa2021}. Here, we propose that the observed motion and alignment in the kHz frequency regime are instead mediated by electrohydrodynamic (EHD) flows. In the low kHz regime (0.5-10 kHz) EHD phenomena dominate when $\rm \kappa D/H \ll f \ll \kappa^2 D$ \cite{Ristenpart2007} (being $\kappa$ the inverse of the Debye length, D the diffusion of ions in solution and H half of the distance between two electrodes). For $\rm SiO_2$ colloidal spheres in water, these EHD flows are attractive, with flow profiles directed toward the particle volume \cite{Ma2015a, Ma2015}. Thus, at $\rm f < 2kHz$ when two $\rm SiO_2$-metal Janus particles approach each other with their $\rm SiO_2$ heads in front, the attractive EHD flows around the $\rm SiO_2$ hemisphere induce short-range attraction, causing the particles to remain together rather than immediately scattering. Following a collision, the particles tend to align, indicating the presence of a torque, which we speculate could be the result of the hydrodynamic coupling between the fluid flow profiles amongst two particles \cite{llopis2010hydrodynamic, molina2013hydrodynamic}. These interactions, however, are weak, leading to scattering after an alignment time $\rm \tau_a =$ 0.5-2~s (Supplementary Fig. S3). 

In view of capturing the main mechanisms responsible for the experimental observations in simulations, we neglect an explicit hydrodynamic description and adopt a simplified "dry" picture, based on the Active Brownian Particle (ABP) model with alignment interactions \cite{MartinGomez2018, sese2018velocity,  sese2021} (see Methods). In this model, the frequency-dependent interactions between colloids are quantified by the dimensionless parameter  
\begin{equation}
\label{eq:globalP} 
\rm g =  \frac{2J}{D_r \pi R_{\varphi}^2}  ;
\end{equation}
which reflects the strength of alignment as compared to rotational noise, and enables the quantitative analysis of its impact on the collective behaviour. Here $\rm J$ is the alignment coupling strength, $D_r$ the single particle rotational diffusion and $R_{\varphi}$ the alignment interaction range. Thus, one can map the role played by the frequency $\rm f$ in the experiments by varying $\rm g^{-1}$ in the ABP simulations (see Methods). This alignment effect should be further elucidated by detailed numerical studies including full electro-hydrodynamic effects, which we speculate also play an important role in the short-range attraction and particle alignment. 

In the following sections, we will explore the collective behaviour at a fixed voltage of 8 $\rm V_{pp}$ as a function of area fraction and frequencies from 0.8 to 2 kHz, thus investigating cases of strong and moderate alignment and non-alignment, respectively. We will start with a single population and then investigate binary mixtures.

\subsection{Emergent polar cluster phase in monodisperse populations}

\begin{figure}
\includegraphics[width=0.85\linewidth]{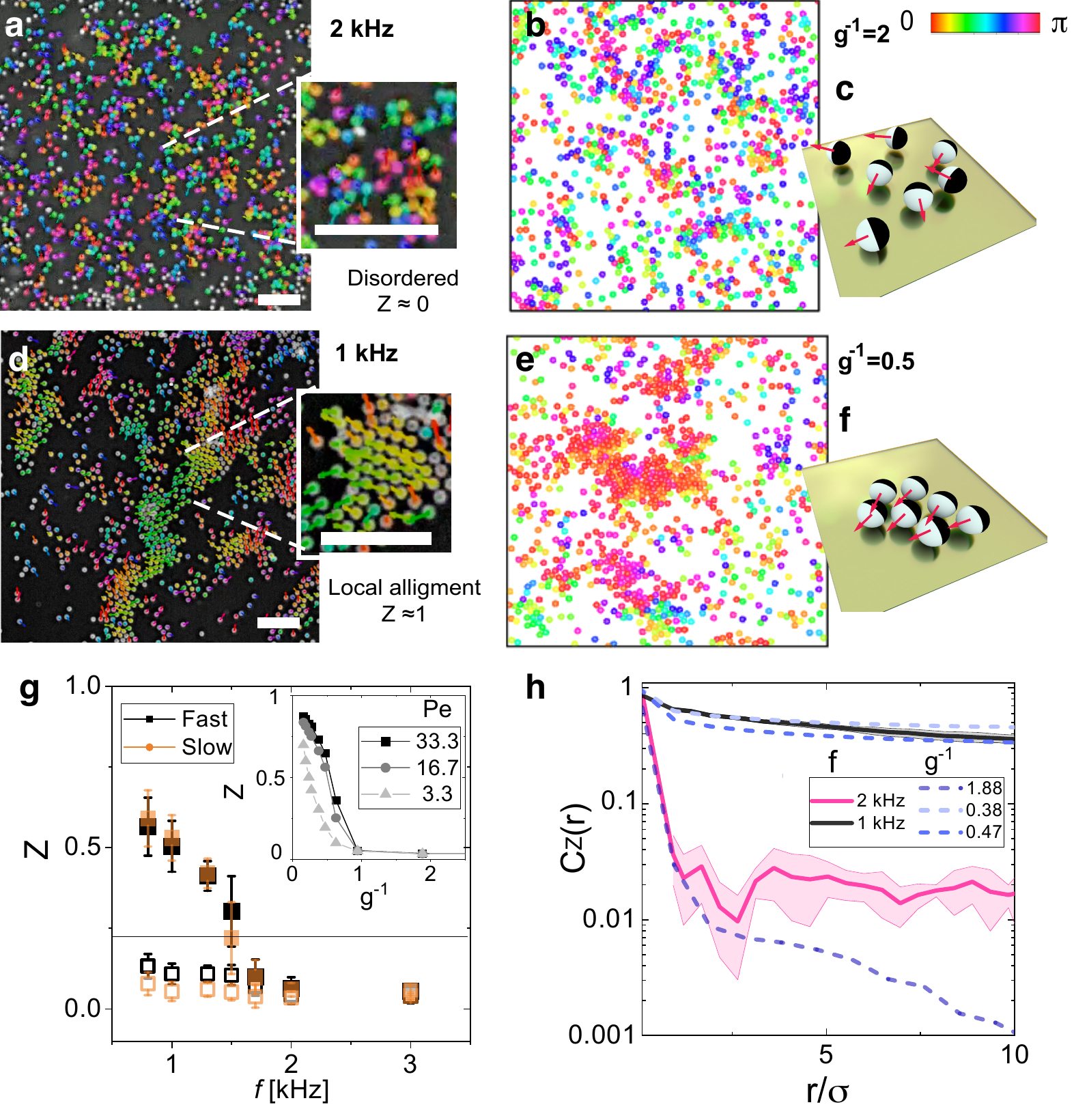}
\caption{\textbf{Emerging collective behavior of monodisperse systems.}~\textbf{a-d}, Snapshots of optical microscopy movies of the fast Janus particles at A\% $\approx$ 0.2 area fraction for 8 $\rm V_{pp}$ and f= 2 kHz~\textbf{a} and 1 kHz~\textbf{d}. The colour arrow code represents the velocity vector and particle orientation $\rm \theta$. The insets are a zoom-in of each corresponding snapshot. The arrows depict the velocity vector. Scale bars indicate 20 $\mu m$.~\textbf{b-e}, Analogue numerical simulation snapshot for weak $\rm g^{-1}$= 2 and strong $\rm g^{-1}$=0.5 alignment.~\textbf{c,f} Corresponding schemes depicting disordered and polar cluster configurations.~\textbf{g} Global polarization $\rm Z$ as a function of frequency for A\%$\approx$ 0.01 (open), and A\%$\approx$ 0.2 (solid) for slow $\rm Pe^{exp}_{slow} \approx 60$ and fast $\rm Pe^{exp}_{fast} \approx 130$ particles (orange and black symbols, respectively). Inset represents the results from numerical simulations at matching values of $\rm g^{-1}$ at  $\rm A\% \approx 0.2 $ for $\rm Pe^{sim}= 3.3$, $\rm Pe^{sim}= 16.7$ and $\rm Pe^{sim}= 33.3$ (from black to light grey). \textbf{h},~Polar correlation $\rm C_Z(r)$ as a function of inter-particle distance normalized by the particle diameter $\rm \sigma$ for experiments at 1 (black) and 2 (magenta) kHz at 0.2 A$\%$, and for numerical simulations (dotted lines) at fixed ${\rm Pe}=16.7$ and the different values of $\rm g^{-1}$ shown in the legend.}
\label{fig:fig2}
\end{figure}

We investigate the effect of the area fraction (A\%) and alignment interactions on particle dynamics, supported by an ABP model with alignment interactions. In Fig.\ref{fig:fig2}a,d we display typical experimental configurations for the fast particles at two different frequencies and fixed voltage for area fraction A\% $\approx$ 0.2. In Fig.\ref{fig:fig2}d, we illustrate the microstructural change the system exhibits while crossing the transition from a disordered state to a polar cluster phase when decreasing the frequency from 2 to 1 kHz (Supplementary  Movie S3 and S4). This behavior is reminiscent of dynamical structures found in model systems of self-propelled particles with alignment interactions in a relatively dilute disordered background \cite{chate2008collective, peruani2012collective, MartinGomez2018, chate2020dry}. The same phenomena is nicely captured in the ABP model simulations (Fig.\ref{fig:fig2}b,e and Supplementary Movie S5), where increasing the alignment parameter $\rm g$ results in a transition from a disordered state to polar clusters, eventually leading to a large (macroscopic) high-density polar structure \cite{MartinGomez2018, sese2021}. 

The transition from a disordered to a polar cluster phase can be quantified by computing the global polarisation $Z$ as an order parameter to describe the system's state. It is defined as
\begin{equation}
\label{eq:globalP} 
Z = \norm {\frac{1}{N}  \sum_{i}^{N} \textbf{n}_i} ;
\end{equation}
with $\rm n_i=v_i/|v_i|$ as in Vicsek-like models \cite{Vicsek1995, chate2008collective, MartinGomez2018}. 
In Fig.\ref{fig:fig2}g we show the typical values of $\rm Z$ for a frequency range from 0.8 to 3 kHz at a fixed voltage value of 8 $\rm V_{pp}$, for area fractions A$\%$ $\approx$ 0.01 and  A$\%$ $\approx$ 0.2 of fast $\rm SiO_2$-Pd and slow $\rm SiO_2$-Au Janus colloids independently (see Fig.\ref{fig:fig2}c). 

We observe that the global alignment parameter $\rm Z$ strongly depends on particle density and frequency. At very low area fractions (A$\% < 0.01$), $\rm Z$ increases slightly to $\approx 0.1$ as the frequency decreases from 1.5 kHz, reflecting rare alignment events between pairs of particles. At intermediate densities (A\% $\approx 0.2$), the system is disordered at high frequencies ($\rm f > 1.5$ kHz) with $\rm Z < 0.25$, but, below 1.5 kHz, $\rm Z$ increases significantly ($\rm Z > 0.25$), corresponding to the emergence of polar clusters. As the frequency decreases further, $\rm Z$ reaches values as high as $\approx 0.6$ for both fast and slow particles, driven by finite reorientation times and the dynamical formation-rupture of clusters. Although individual clusters exhibit high local order ($\rm Z \approx 1$, Supplementary Fig.S4), the global alignment parameter averages to $\rm Z \approx 0.6$ due to variations in cluster orientations across the system in this parameter regime. 
These findings are supported by our ABP simulations at A\% $\approx 0.2$ over a range of $\rm g^{-1}$ values and Péclet numbers (3.3, 16.7, and 33.3). Remarkably, the simulations show qualitative agreement with experiments, confirming that the onset of polar clusters is primarily governed by particle density and alignment strength ($\rm g^{-1}$), with minimal influence from the Péclet number, in agreement with mean-field predictions and particle-based simulations \cite{Farrel2012, MartinGomez2018}. The difference between the increase of $\rm Z$ in experiments and simulations arises from the limited experimental access to low-frequency regimes. 

We furthermore quantify the effect of the polar clusters formation on particle dynamics by measuring the polarization auto-correlation function $\rm C_z=N^{-1}\sum_i\langle \boldsymbol{n}(r_i)\cdot \boldsymbol{n}(r_i+r)\rangle$  as a function of inter-particle distance $\rm r$ (Fig.\ref{fig:fig2}h) ($\rm \boldsymbol{n}(r_i)$ being the polarization of a particle located in $\rm r_i$). In the gas-like regime, where particles do not align, the polarization correlation decays rapidly and exponentially. At lower frequencies, alignment emerges and polar clusters form, leading to persistent collective motion with strongly correlated particle orientations. This behaviour is similar for both fast and slow particles, reflecting comparable effective attractions from electrohydrodynamic (EHD) flows around the $\rm SiO_2$ hemisphere. Simulations of active Brownian particles capture the same transition, with the orientational correlation shifting from rapid decay to long-range persistence. This allows for a direct mapping between alignment strength in simulations and the AC field frequency in experiments.  

\begin{figure}
\includegraphics[width=1\linewidth]{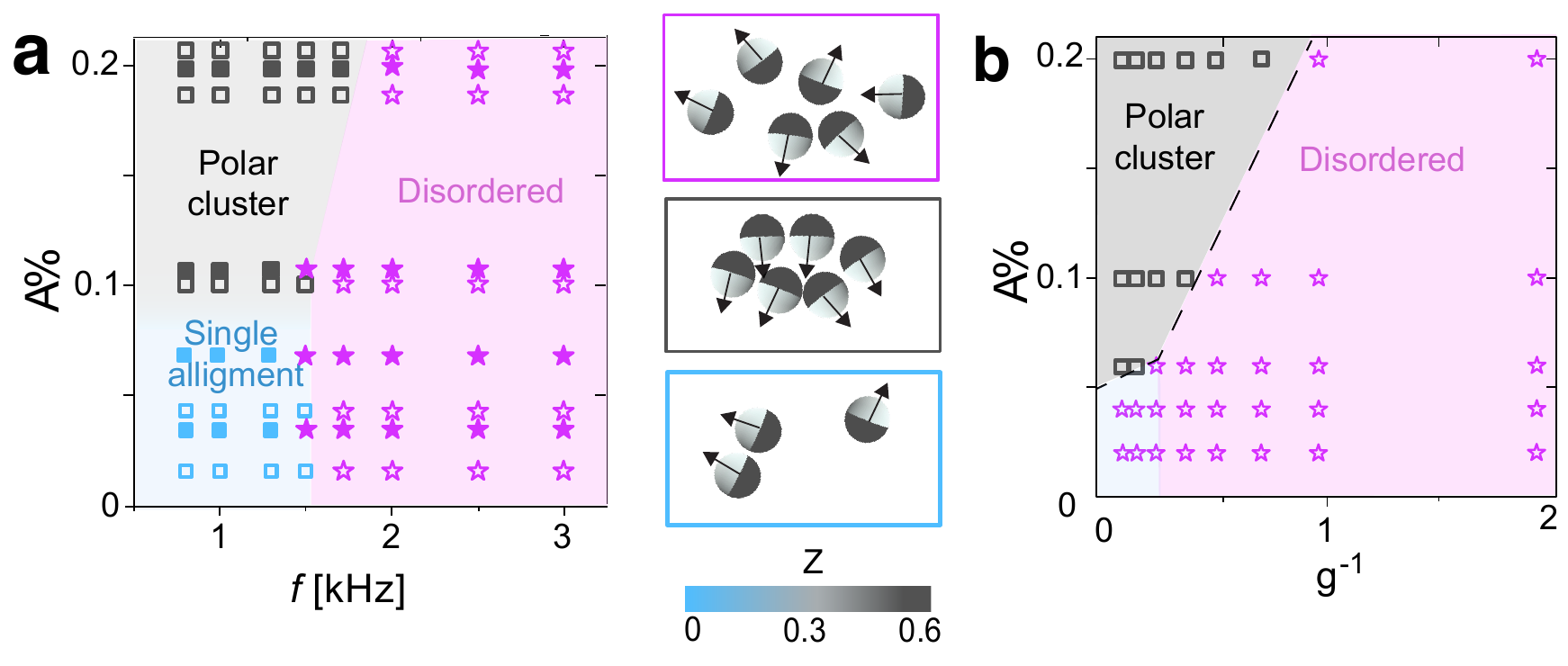}
\caption{\textbf{Phase diagram of fast and slow Janus colloids}~\textbf{a,}~Experimental phase diagram of the collective behaviour of fast $\rm SiO_2$-Pd (open symbols) and slow $\rm SiO_2$-Au (solid symbols) particles as a function of area fraction $\rm A \%$ and applied frequency f (kHz) at a fixed voltage of 8 $\rm V_{pp}$ for fast particles ($\rm Pe^{sim}_{slow}$)~\textbf{b,}~Phase diagram obtained from numerical simulations at \textbf{$\rm Pe= 16, 33.3$}, varying $\rm A \%$ and alignment $\rm g^{-1}$. The schemes show the various states for disordered (magenta), polar cluster (gray) and single alignment (blue) events, corresponding to each phase of the diagram.}
\label{fig:fig3}
\end{figure}

We construct a phase diagram mapping the behaviour of monodisperse suspensions of fast and slow particles as a function of area fraction ($\rm A$) and frequency, at fixed voltage (8~$\rm V_{pp}$) (Fig.~\ref{fig:fig3}). At low frequencies ($f < 1.5$~kHz) and dilute conditions ($A < 0.5\%$), only transient alignment events occur, with particles swimming together briefly before scattering, resulting in a disordered, isotropic phase. At higher frequencies ($f > 1.5$~kHz), alignment is suppressed, and the disordered phase persists across all densities (pink region in Fig.~\ref{fig:fig3}a). As $A$ increases beyond $\approx$ 0.1 and $f <$ 2~kHz, polar clusters emerge with $Z >$ 0.25, characterized by dynamic formation and breakup, reminiscent of bacterial swarming~\cite{Zhang2010, peruani2012collective}. This transition is identified when particles have at least three neighbours within the alignment time and cut-off distance. At even higher densities ($A >$ 0.25\%), vortex-like structures form (Supplementary Fig.S5), as reported in similar systems~\cite{Zhang2020, Zhang2021}. ABP simulations show comparable behaviour (Fig.~\ref{fig:fig3}e), with transitions from disorder to polar clustering governed by $\rm g^{-1}$ and $\rm A$. As in experiments and predicted theoretically~\cite{Farrel2012, MartinGomez2018, chate2020dry}, increasing density enhances alignment for a given coupling strength.

Finally, we also quantify the swimming speed of the slow and fast particles using the distribution of displacements as a function for low and dense A\% (Supplementary Fig.S6). As expected, we see a decrease in the mobility in the dense phase with respect to the diluted phase, mainly induced by crowding effects  \cite{Farrel2012}. 

\begin{figure}[h]
\includegraphics[width=1\linewidth]{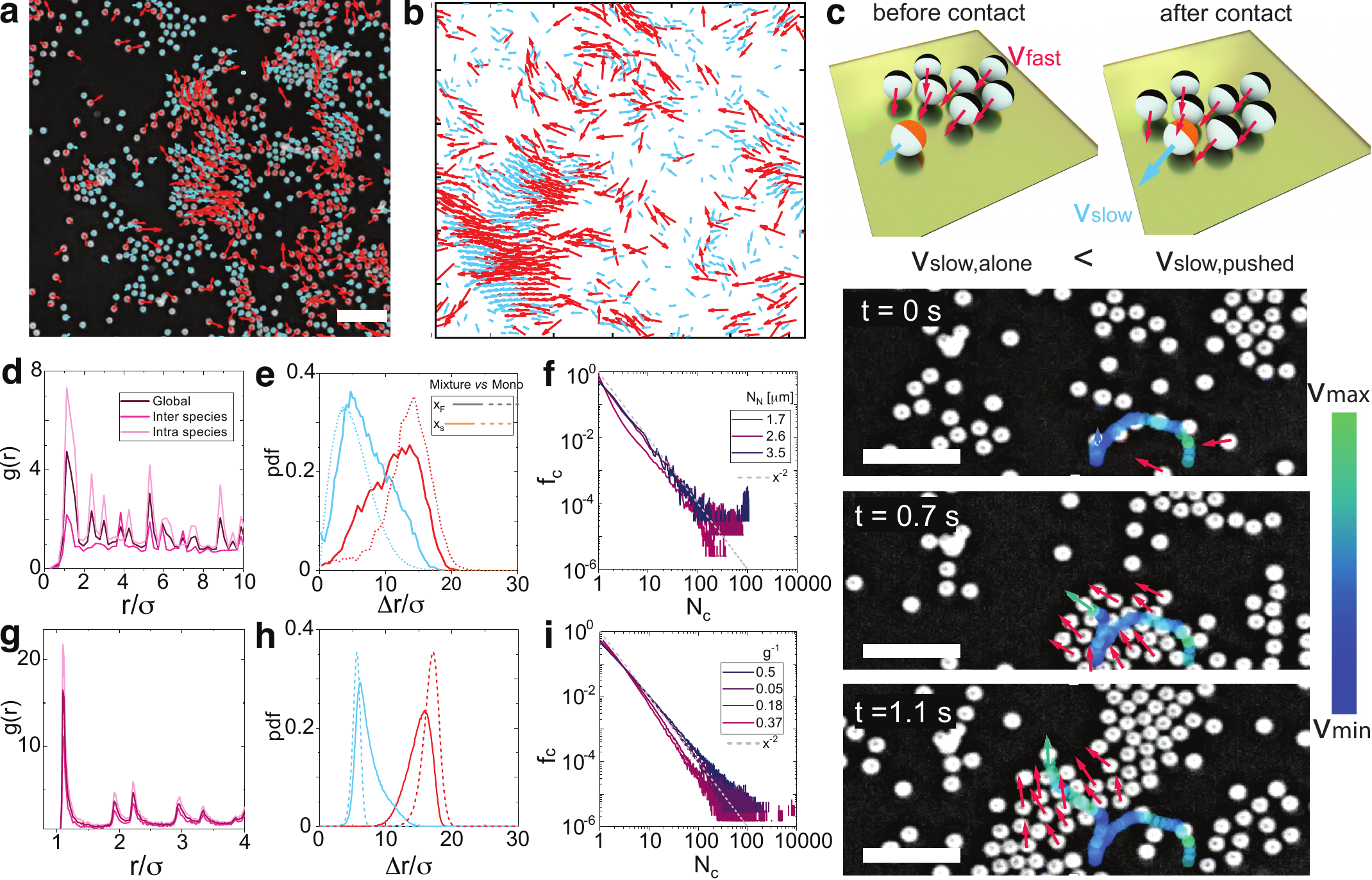}
\caption{\textbf{Emergent segregation and cooperation in binary mixtures.} 
Screenshots from \textbf{a}, experiments at 8~$\rm V_{pp}$, 1~kHz; 
\textbf{b}, and simulations with $\rm g^{-1}=0.18$, $\rm Pe^{sim}_{slow}=3.3$, $\rm Pe^{sim}_{fast}=13$. Mixtures of slow (cyan) and fast (red) particles at area fraction $A\% \approx 0.2$, with $\rm x_{F,exp}\approx 0.45$ and $\rm x_{F,sim}\approx 0.5$. Scale bars: 20~$\rm \mu m$. 
\textbf{c}, Schematic (top) and screenshots from fluorescence movies (bottom) showing a slow particle being pushed by a fast-particle cluster, resulting in enhanced velocity of the slow particle. The arrow depicts the velocity vector for fast (red) and slow (cyan). The trajectory colour represents the instantaneous velocity. \textbf{d,g}, Displacement distributions at $t=0.3$~s from experiments (top) and simulations (bottom) for slow ($\rm x_S$, cyan) and fast ($\rm x_F$, red) particles in mixtures (solid) and monodisperse controls (dashed), $\Delta t=0.1$~s. \textbf{e,h}, Pair correlation function $g(r)$ for intra-, inter-species, and total pairs (light to dark magenta) for experiments (top) and simulations (bottom). 
\textbf{f,i}, Cluster size distributions $\rm f_c(N_c)$ from experiments (top) (cut-offs $\rm N_N=1.7, 2.6, 3.5~\mu$m) and simulations (bottom) varying $\rm g$, using as a cut-off distance to define a cluster the first peak of $g(r)$  \textbf{g}. Dashed line: $\rm f_c \sim N_c^{-2}$}
\label{fig:fig4}
\end{figure}

\subsection{Segregation and species cooperation in binary mixtures}

We further investigated the collective dynamics of dense mixtures of fast $\rm SiO_2$-Pd and slow $\rm SiO_2$-Au Janus colloids, prepared at an approximate 50:50 population ratio. Based on the results from monodisperse suspensions, we selected experimental conditions at 1 kHz and 8 $\rm V_{pp}$ and area fractions, where polar dynamical clusters dominate. As both particles are indistinguishable under optical microscopy, we performed a dynamical classification of the particles based on their individual distribution of displacements (Supplementary Information). Similarly, we compute the analogue scenario with the ABP model in the presence of the fast ($\rm Pe^{sim}_{fast}=$13) and slow ($\rm Pe^{sim}_{slow}=$3) species (within the range that qualitatively agrees with our experiments). In Fig.\ref{fig:fig4}a,b we show representative snapshots of experimental and numerical active binary mixtures comprising fast $\rm SiO_2$-Pd and slow $\rm SiO_2$-Au Janus colloids (Movies S6 and S7, respectively). 

In contrast with previous works using catalytic or light-driven Janus particles \cite{Kummel2015, Singh2017}, slow particles do not necessarily get trapped in clusters of fast ones. In fact, a spatial segregation between the two species is observed. Such organization can be evidenced from the analysis of the pair correlation function $\rm g(r)$, as shown in Fig.\ref{fig:fig4}d,g. We compute separately the pair-correlation functions considering only correlations between particles of the same species -- being fast/fast or slow/slow --, between particles of opposite species (fast-slow) and all particles without distinction between the species. Here we observe that particles of the same species are more likely to cluster together, evidenced by higher and sharper peaks in the  $\rm g(r)$.

\begin{figure}
\includegraphics[width=0.5\linewidth]{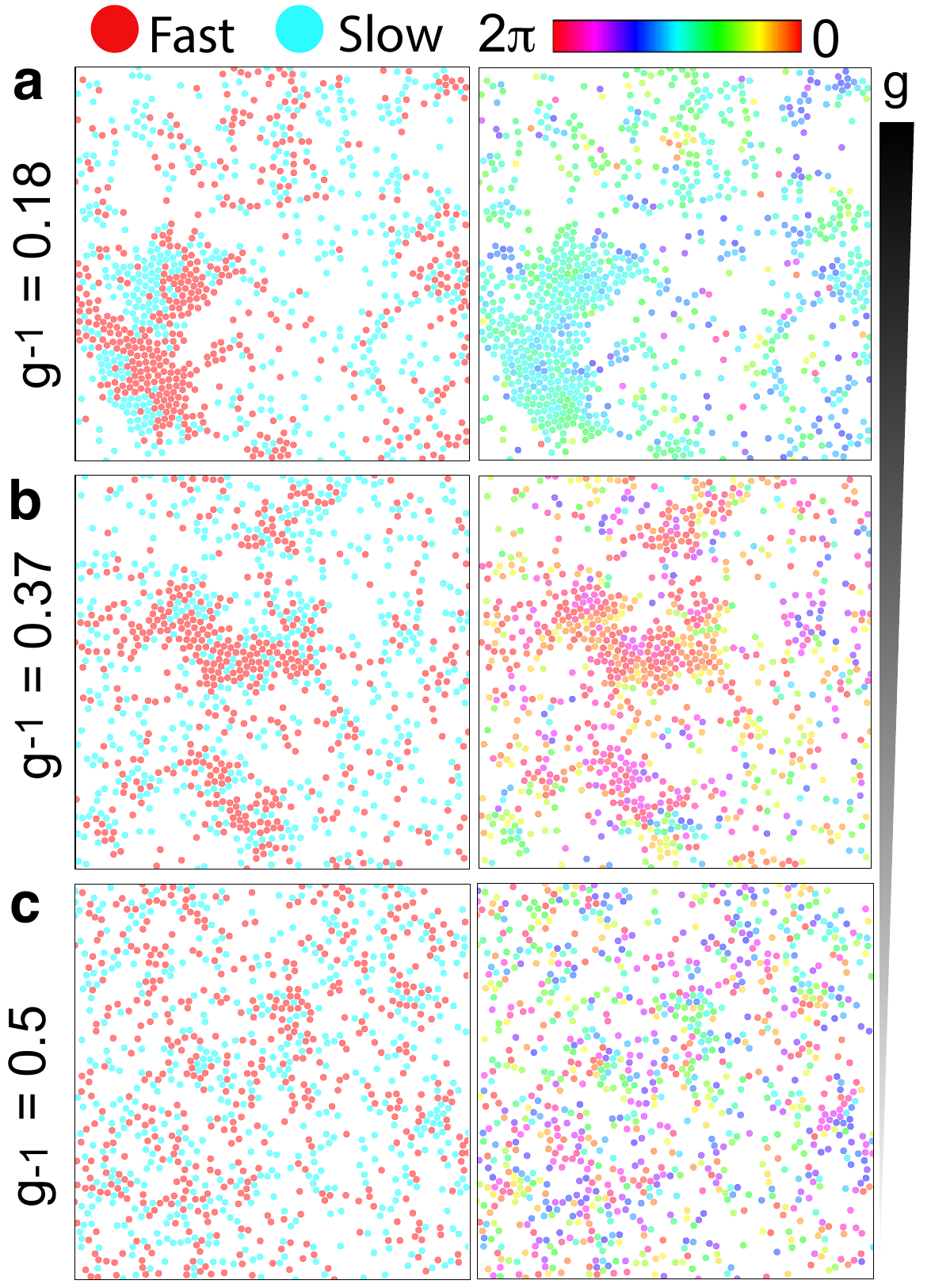}
\caption{\textbf{Simulation of binary mixtures of active particles as a function of alignment strength}. Snapshots from simulations with two species with different alignment strength ~\textbf{a} $\rm g^{-1} = 0.18$,~\textbf{b} $\rm g^{-1} = 0.37$, and ~\textbf{c} $\rm g^{-1} = 0.5$. The fast particles ($\rm Pe^{sim}_{fast}=13$) are in red and the slow ones ($\rm Pe^{sim}_{slow}=3$) in cyan. (left column).  The graphs in the right column show the corresponding particle orientations. In all the simulations $\rm x_F = 0.5$ and A\% $\approx 0.2$.}
\label{fig:fig5}
\end{figure}

In addition, the polar clusters formed by each species interact with each other, where fast particles tend to push the polar clusters of slow particles before the clusters break. These observations are further quantified by examining the distributions of displacements. As shown in Fig. \ref{fig:fig4}e,h for experiments and simulations, the distributions of displacements for mixtures  -- with fast and slow particles --  with area fraction A\% $\approx$ 0.2 where the $\rm x_{F} \approx$ 0.5, exhibit distinct tails that are not present in the monodisperse case, and the overall displacements for the slow particles are slightly higher, while they are slower for the fast ones. In particular, for slow particles, we observe a residual tail shifted towards higher displacements, with slow particles moving faster in the bidisperse mixture than they would in the corresponding monodisperse case. Likewise, fast particles exhibit a tail towards lower values. We attribute this effect to the interaction between the two species when they form polar clusters.  In our simulations, this effect is even more pronounced (Supplementary Movie S7). Such behavior is illustrated in the scheme in Fig.\ref{fig:fig4}c and Supplementary Movie S8, where a slow particle is pushed by a fast particle cluster, with its instantaneous velocity increasing while being dragged by the fast species.
 
\begin{figure}
\includegraphics[width=0.4\linewidth]{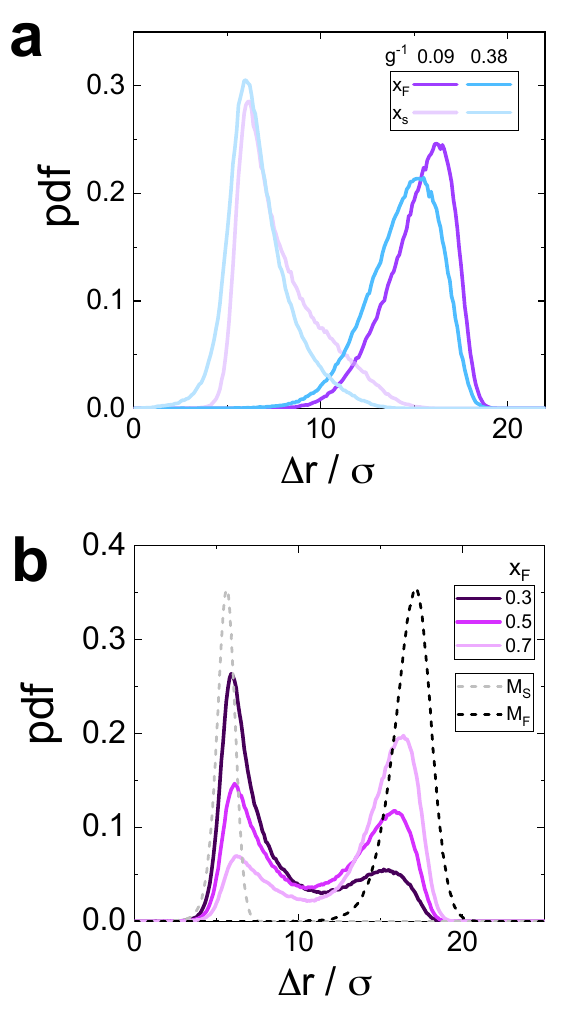 }
\caption{\textbf{Effect of particle ratio and alignment in mixtures.} PDF of displacements from simulations with dt = 0.1$\rm s$ in a binary mixture of slow and fast particles ($\rm Pe^{sim}_{fast}=13$, $\rm Pe^{sim}_{slow}=3$), and A $\approx$ 20 $\%$ total particle density for: ~\textbf{a} A varying fraction of fast particles at fixed alignment $\rm g^{-1}=0.18$ and varying species ratio $\rm x_F = 0.3$ (dark magenta) $\rm x_F = 0.5$ (magenta), and  $\rm x_F = 0.7$ (light magenta). The dashed lines represent the monodisperse case for slow (transparent) and fast (solid) particles. ~\textbf{b} A fixed particle ratio $\rm x_F = 0.5$, and varying the alignment force $\rm g^{-1}$. The black and gray line indicates the fast species and slow species, respectively.}
\label{fig:fig6}
\end{figure}
 
Thus, particles of the same species are more likely to cluster together, a signature of the structure described earlier: dense swarms mainly composed of fast particles push slow particles from the back, generating segregation of fast and slow particles within the dense polar structures. Note that such morphology has a purely dynamical origin, as the onset of polar order is identical for slow and fast particles (see Fig. \ref{fig:fig2}a). The faster-moving polar swarms composed of Pd colloids drag the (also polar) swarms of slower Au colloids. We measure the normalized frequency of cluster size $\rm f_c$ at various cluster sizes (Fig.\ref{fig:fig4}f,i). We observe a rather broad distribution of cluster sizes,  with a decay of $\rm \propto  N_c^{-2}$, meaning that the system is populated by clusters of any size, with no typical length scale. 

As it is experimentally not possible to control the initial location of slow and fast particles, nor the in-out flux of both species within the region of interest while acquiring the videos, systematically studying how the ratio of species influences the cooperative behaviour is beyond our experimental capabilities. Conversely, we can fully explore the effect of species ratio using numerical simulations for different values of $\rm g^{-1}$. The latter is crucial for the segregation of both phases and the positioning of the population with respect to each other. In Fig.\ref{fig:fig5}, we show quantitatively the effect of alignment strength in binary mixtures (Supplementary Movie S9). Here, we always observe the emergence of a polar cluster phase - with high local ordering -  above a certain threshold of alignment, where the fast particles always push groups of slow particles, which are eventually left behind or disassembled. This is in stark contrast with previous experimental and numerical works done in the regime of motility-induced phase separation, where slow or passive particles get trapped within clusters of the fast ones. 

Moreover, we measured the distribution of displacements fixing the fraction of slow and fast particles ($\rm x_F =$ 0.5) at a total area fraction of A$\%$ $\approx$ 0.2. The effect of alignment strength seems to affect mostly the slowest particles with residual tails shifting towards higher displacements (Fig.\ref{fig:fig6}a). Most importantly, the ratio between each species determines the overall behaviour of the mixture, based on the inter-species interaction, and thus their relative displacements (Fig.\ref{fig:fig6}b). Not surprisingly, we see an overall effect on the mixture led by the dominant population: when the majority of particles are fast, the slow particles tend to become faster, and vice-versa. Not only is there a slight shift on the characteristic peak of the distribution of displacements affecting their relative speeds, but the distribution becomes biased and exhibits a valley that shows intermediate behaviour, as previously shown when compared to the experimental conditions. This is not present in the monodisperse case. Thus changing $\rm g^{-1}$, the mean peak position of the displacements can be shifted (the motility of each species is affected), while the effect of the fraction of the populations controls the weight of each population, where one species dominates over the other one).  

\section*{Conclusions}

Here we demonstrate how to engineer binary mixtures of active colloids by exploiting the interplay between particle material properties and external AC electric fields. We fabricate two types of Janus particles with distinct motilities yet similar rotational diffusion, enabling precise control over their alignment and interactions as a function of field frequency. Our experiments show that emergent polar clustering can be selectively induced by adjusting the frequency of the applied electric field. 
 
Remarkably, we control alignment and collective effects in a very small range of frequencies in the kHz regime, relative to previous work with Janus colloids \cite{Yan2016, Zhang2021,Iwasawa2021}, and at notably lower applied voltages than, for instance, in Quincke rollers set-ups \cite{Bricard2013}. These conditions allow for fine-tuning experimental parameters and provide an opportunity to explore collective effects in mixtures of self-propelled particles. Our results reveal the two species of particles can self-organize into polar clusters of fast- and slow-moving species. At high densities, the interactions between fast and slow particles lead to effective segregation and cooperative dynamics, whereas fast particles lead to an effective speed-up of slower ones. Our experimental observations are supported by numerical simulations using a simple model of active Brownian particles with varying alignment strength, which effectively captures the essential features of the observed behaviour without explicit electro-hydrodynamic effects. The model provides a framework to map the role of alignment strength and particle density, leading to a deeper understanding of the conditions under which collective behaviour emerges. As this is purely material-dependent \cite{Yang2019}, the swimming behaviour and potential interactions can be tuned using different dielectric materials and metals to fabricate and design rich collective phenomena in Janus particle mixtures.
Finally, the ability to fine-tune these interactions opens up new possibilities for designing reconfigurable active materials. Such control over particle behavior and interactions could be interesting for active materials with mechanical properties that depend on their internal structure, as previously proposed in active \textit{fractalytes} \cite{Fehlinger2023}. Moreover, the intrinsic differences in motility and flow fields around the particles could be exploited as two-species model systems with non-reciprocal interactions \cite{dinelli2023non}.

\section*{Acknowledgements}

L.A. thanks Caroline van Baalen for her support in sputter coating the particles, and Federico Paratore for the $\rm SiO_2$ coating of conductive slides. L. A. thanks Nicolas Bain for fruitful discussions. L.A. thanks SNFS (Spark grant No. 190392), IdEx Bordeaux and the French Agence Nationale
de la Recherche (ANR-23-CE06-0007-0) for funding. D.L. acknowledges MCIU/AEI and DURSI for financial support under Projects No. PID2022-140407NB-C22 and 2021SGR-673, respectively. I.P. acknowledges support from
Ministerio de Ciencia, Innovaci\'on y Universidades
MCIU/AEI/FEDER for financial support under grant
agreement PID2021-126570NB-100 AEI/FEDER-EU, and  Generalitat de Catalunya for financial
support under Program Icrea Acad\`emia and project
2021SGR-673. L.I. acknowledges funding from the European Research Council (ERC) under the European Union’s Horizon 2020 Research and Innovation Program grant agreement No 101001514.

\section*{Author Contributions}
Author contributions are defined based on the CRediT (Contributor Roles Taxonomy) and listed alphabetically. Conceptualization: L.A., L.I., D.L., I.P.;  Formal analysis: L.A, D.L., E.S.S.,  Funding acquisition: L.A., L.I. Investigation: L.A., D.L., E.S.S.  Methodology: L.A., D.L., E.S.S. Project administration: L.A., D.L.; Software: L.A., D.L., E.S.S. Supervision: L.I., D.L., I.P.; Validation: L.A., D.L., E.SS. Visualization: L.A. Writing – original draft: L.A., D.L., E.S.S. Writing – review, and editing: L.A., D.L., I.L., I.P.

\section*{Additional Information}
Supplementary Information is available for this paper. Correspondence and requests for materials should be addressed to L.A. and D.L. 

\section*{Data availability statement}
The data that support the findings of this study are available from the corresponding authors upon reasonable request.

\section*{Code availability statement}
The code used in this study is available from the corresponding authors upon reasonable request.

\section*{Methods}
\subsection*{Fabrication of Janus particles}
To fabricate our active Janus particles, silica colloids with 2.1 $\rm \mu$m diameter (5\% w/v, microParticles GmbH, Germany) are diluted to 1:6 in MilliQ water and spread on a glass slide, previously made hydrophilic by a 2-minute air plasma treatment. Upon drying the suspension, a close-packed particle monolayer is formed. The monolayer is then sputter-coated (Safematic CCU-010, Switzerland) with 2 nm Cr for both particles, a second layer with 4 nm of Au for one type of particle (slow), and 6.5 nm of Pd for the other particle (fast). The Cr layer is used to promote the adhesion of the other metals to silica. The particles are retrieved from the glass slide by sonicating the slide containing the coated monolayer in MilliQ water. For the bidisperse experiments, we prepare a stock solution using the same volume of the Pt-$SiO_2$ and Au-$SiO_2$ particle in MilliQ water (conductivity $\rm \sigma_m$ = 5x10$^{-5}$ Sm$^{-1}$) .

\subsection*{Cell preparation}
The transparent electrodes are fabricated from 18 mm $\times$ 18 mm No. 0 cover glasses ($85-115\,\mu $m-thick, Menzel Gl\"aser, Germany) covered with 3 nm of Cr and 10 nm of Au, deposited by metal evaporation (Evatec BAK501 LL, Switzerland), followed by 10 nm of SiO$_2$, deposited by plasma enhanced chemical vapour deposition (STS Multiplex CVD, UK) to minimize the adhesion of particles to the substrate during the experiments. A water droplet containing the particles is deposited on the bottom electrode within the 9 mm-circular opening of a 0.12 mm-thick sealing spacer (Grace Bio-Labs SecureSeal, USA). The top and bottom electrodes are connected to a signal generator (National Instruments Agilent 3352X, USA) that applies the AC electric field,  varying frequencies from 0.8 to 5 kHz and varying the voltage between 1 and 10 $V_{\rm PP}$.

\subsection*{AC electric field experiments}
The Janus particles are imaged with a Zeiss Axio microscope in transmission and image sequences are taken with a sCMOS camera (Andor Zyla 4.2) at varying frame rates between 10 and 100 fps depending on the swimming speed and concentration of the Janus particles, with a variable field of view among experiments. The positions of the centre of the JPs and the metal cap are located using Matlab routines. Then, a vector connecting both centres is used to determine the orientation of the particle at each frame. In addition, the centre of mass of the particle is used to determine the velocity vector. The image series of the metallic Janus particles actuated by the AC electric fields are acquired with a 20$\times$ and 40$\times$ objective (Zeiss). 

\subsection*{Particle-based model}

We consider a system of self-propelled particles in a two-dimensional squared box of side $L$ with periodic boundary conditions (PBC). The equations governing the dynamics of the system are the coupled over-damped Langevin equations,

\begin{equation}
\dot{\textbf{r}}_{i} = v_{0}^{\alpha} \textbf{n}_{i} + \mu \textbf{F}_{i}+ \sqrt{2D_0}\boldsymbol{\eta}_{i},
\label{eq:eomR-mix}
\end{equation}
\begin{equation}
\dot{\varphi}_{i} = J \sum_{j\neq i} v(r_{ij})\sin(\varphi_{j}-\varphi_{i}) + \sqrt{2D_{r}}\nu_{i} .
\label{eq:eomT-mix}
\end{equation}
Each particle $i \in \{1,..., N \}$ self-propels at constant speed, $v_0^{\alpha} \in \{v_0^F,v_0^S \}$, where $v_0^F$ is the nominal speed of the fast species, and $v_0^S$ of the slow one. 
$\boldsymbol{\eta}_{i}$ and $\nu_{i} $ are two independent white Gaussian noises with zero mean and unit variance: $\langle {\eta}_{i}^{\alpha}(t) {\eta}^{\beta}_{j}(t^{\prime}) \rangle=\delta_{ij}\delta^{\alpha \beta}\delta(t-t^{\prime})$ and  $\langle {\nu}_{i}(t) {\nu}_{j}(t^{\prime})\rangle=\delta_{ij}\delta(t-t^{\prime})$ ($\alpha$, $\beta$ denoting cartesian coordinates). Rotational noise sets a characteristic time scale, given by the inverse of the rotational diffusion coefficient, $\tau = D_r ^{-1}$, the persistence time, and a characteristic length scale given by $l_p=v_0\tau$, the persistence length. The noise $\boldsymbol{\eta}_{i}$ mimics a thermal bath surrounding the particles at temperature $T$. The diffusivity $D_0$ and mobility $\mu$  fulfill the Einstein relation $D_0=\mu k_B T$.

Steric interparticle forces, $\textbf{F}_{i} = - \sum_{j \neq i} \nabla_i u \left(r_{ij} \right)$, derive from a WCA potential,
\begin{equation}
u (r_{ij}) =\begin{cases} 4 u_0 \left[ (\frac{d}{r_{ij}})^{12} - (\frac{d}{r_{ij}})^6 \right] + u_0 & r_{ij} \leq R \\
0 & r_{ij} > R,
\end{cases}
\label{modelmeth_eq:3}
\end{equation}
where $r_{ij}=|\bold{r}_i-\bold{r}_j|$. The cutoff distance is $R=2^{1/6} d$ and $u_0$ corresponds to the characteristic energy scale of the potential.

Aligning torques are introduced in terms of the sum of the phase difference, as in models of ferromagnets, which runs over the particle's vicinity, defined by a spatially decaying function
 \begin{equation}
v(r_{ij}) = \begin{cases} \frac{2}{\pi R^2_{\varphi}} (r_{ij}-R_{\varphi})^2 &  r_{ij} < R_{\varphi}\\ 
0 &r_{ij} > R_{\varphi},
\end{cases}
\label{modelmeth_eq:5}
\end{equation}
where the cutoff distance $R_{\varphi}$ sets the interaction range beyond which particles do not align.
The strength of the aligning torques is controlled by the coupling parameter $J\geq0$, which mimics an alignment mechanism of non-steric origin, in the sense that it is decoupled from the particles' shape.

\subsection{Brownian dynamics simulations}

We choose the particles' diameter $d =1$ to be the unit of length, $\tau_0 = \frac{d^2}{D_0} =1$ the unit of time and $k_BT=1$ the unit of energy.
The strength of the pairwise repulsive interaction is $u_0 = 100 k_BT$. The thermal and rotational diffusion coefficients fulfil $D_0=\frac{d^2D_r}{3}$, which leads to a fixed rotational diffusion coefficient $D_r = 3/\tau_{0}$. The alignment cut-off distance is set to $R_{\varphi} = 2 d$ and therefore the particles need to be in contact to align their directions of self-propulsion.
We identify the dimensionless parameter $g=\frac{2J}{D_r\pi R_{\varphi}^2}$, which accounts for the strength of the alignment interaction compared to the rotational diffusion and the P'eclet number, $\text{Pe} =\frac{v_0}{d D_r}=\frac{l_p}{d}$, which quantifies the persistence of the particle's motion. The global packing fraction is $\phi=\frac{N \pi d^2}{4 L^2}$. 

In binary mixtures, the ratio between the fast and the slow self-propulsion speed is defined as $v_0^F = f_v v_0^S$. In addition to the speed of self-propulsion, the rest of the parameters are identical for both species. The fraction of fast particles is $x_F=N_F/N$ and the fraction of slow ones is, correspondingly, $x_S=1-x_F$. We thus have a parameter space comprising $(\phi,g,\text{Pe}^S,f_v,x_F)$, where $\text{Pe}^S$ stands for the P\'eclet number of the slow species. 

The Langevin equations defining the model are integrated using an Euler-Maruyama scheme, employing a timestep $\Delta t = 10^{-4}$. Results are computed by averaging over $1000$ independent configurations in the stationary state.

\bibliography{references}

\end{document}



\title{Supporting Information for: Segregation and cooperation in active colloidal binary mixtures}

\author{Laura Alvarez}
\email{laura.alvarez-frances@u-bordeaux.fr}
\affiliation{Univ. Bordeaux, CNRS, CRPP, UMR 5031, F-33600 Pessac, France}%
\affiliation{Laboratory for Soft Materials and Interfaces, Department of Materials, ETH Zurich, 8093 Zurich, Switzerland}%
\author{Elena Ses\'e-Sansa}%
\affiliation{American Physical Society, 100 Mtr Pkwy, Hauppauge, New York 11788 (USA)}%
\author{Demian Levis}
\email{levis@ub.edu}
\affiliation{Computing and Understanding Collective Action (CUCA) Lab, Condensed Matter Physics Department, Universitat de Barcelona, Mart\'i i Franqu\`es 1, E08028 Barcelona, Spain}%
\affiliation{University of Barcelona Institute of Complex Systems (UBICS), Mart\'i i Franqu\`es 1, E08028 Barcelona, Spain}
\author{Ignacio Pagonabarraga}
\affiliation{Condensed Matter Physics Department, Facultat de Física, Universitat de Barcelona, Mart\'i i Franqu\`es 1, E08028 Barcelona, Spain}%
\affiliation{University of Barcelona Institute of Complex Systems (UBICS), Mart\'i i Franqu\`es 1, E08028 Barcelona, Spain}
\author{Lucio Isa}%
\affiliation{Laboratory for Soft Materials and Interfaces, Department of Materials, ETH Zurich, 8093 Zurich, Switzerland}

\maketitle

\onecolumngrid
\parindent 0mm
\bigskip

\section*{Supplementary Movies}

\textbf{Movie S1}. Diluted phase (A$\%$=0.1) of active Janus particles at 1 kHz and 8$\rm V_{pp}$.
\\[4mm]
\textbf{Movie S2}. Alignment events in a monodisperse diluted sample of Janus particles at 1 kHz and 8$\rm V_{pp}$. 
\\[4mm]
\textbf{Movie S3}. Experimental movie of a semi-dense phase (A\% $\rm \approx 0.15$) of fast particles transitioning from elastic collisions (2 kHz, $\rm g^{-1}>1$) to polar alignment (1 kHz, $\rm g^{-1}<1$) at a fixed voltage 8 $V_{pp}$.
\\[4mm]
\textbf{Movie S4}. Experimental dense phase (A\% $\rm \approx 0.2$) of fast particles at the disorder (2 kHz) and polar clusters (1 kHz) phase with their respective velocity vectors at 8$\rm V_{pp}$. 
\\[4mm]
\textbf{Movie S5}. Numerical movie of dense phase (A\% $\rm \approx 0.2$) for fixed alignment $\rm g^{-1}=$0.5 at varying Pe. 
\\[4mm]
\textbf{Movie S6}. Experimental dense binary mixtures (A\% $\rm \approx 0.2$) of slow (blue) and fast (red) ($\rm x_{F} \approx 0.47$) at 1 kHz and 8$\rm V_{pp}$.
\\[4mm]
\textbf{Movie S7}. Numerical dense phase of binary mixtures of slow (blue, $\rm Pe^{sim}_{slow}= 3.3$) and fast (red, $\rm Pe^{sim}_{slow}= 13.3$) particles ($\rm x_{F} \approx 0.5$), decreasing the alignment force $\rm g^{-1}$.
\\[4mm]
\textbf{Movie S8}. Movie of an event of a single slow particle being pushed by a fast cluster. The coloured trajectory depicts the instantaneous velocity of the slow particle (blue) at 1 kHz and 8$\rm V_{pp}$.
\\[4mm]
\textbf{Movie S9}. Numerical simulations movies for binary mixtures at different alignment force $\rm g^{-1}$ at A\% $\rm \approx 0.2$.

\parindent 4mm  
   
\newpage

\begin{figure}[H]
\centering
\includegraphics[width=0.8\columnwidth]{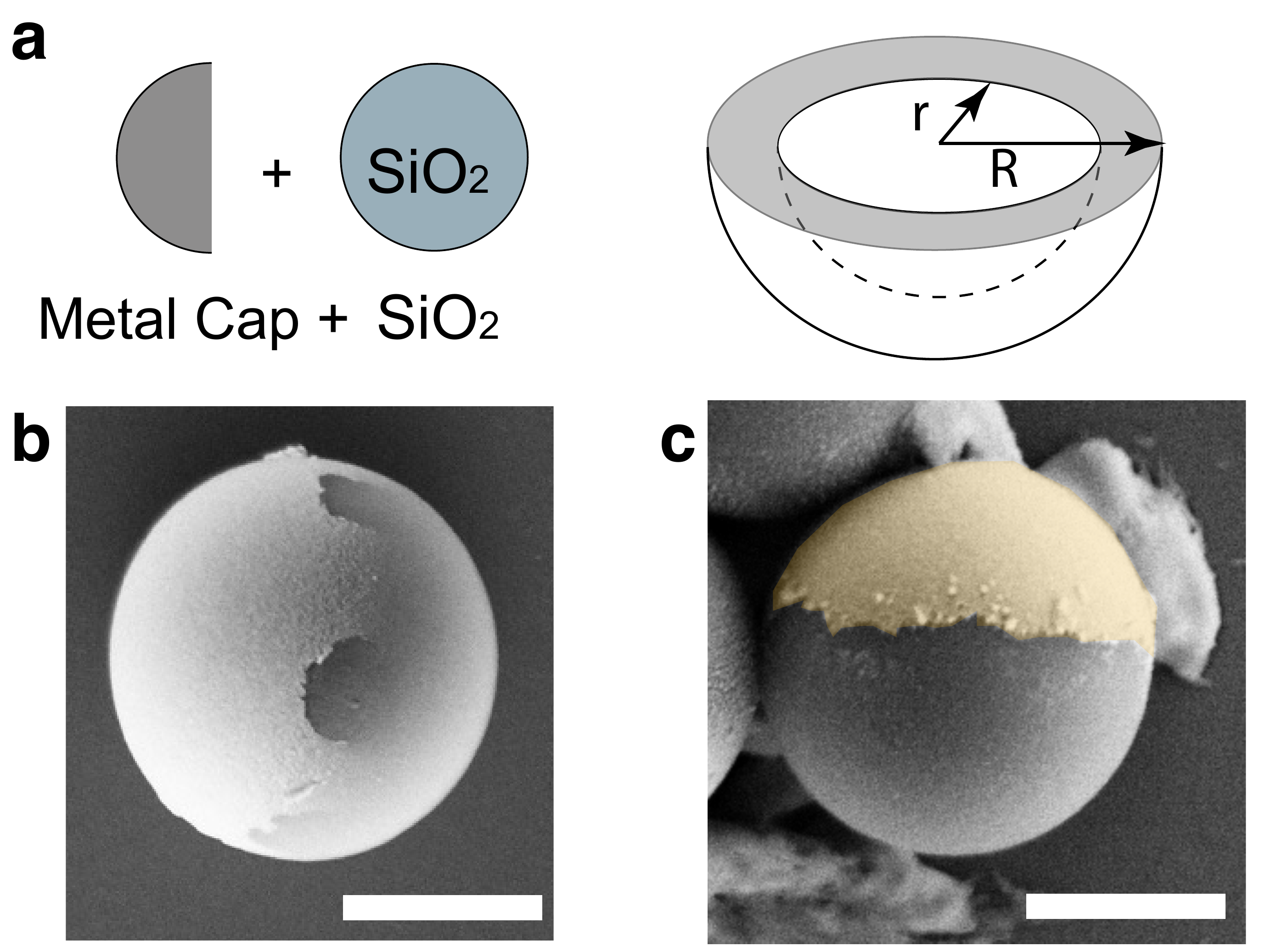}
\caption{\textbf{a}, Scheme of the Janus sphere composed by a metal cap and a $\rm SiO_2$ particle, with the corresponding scheme of the cap, being r the inner radius and R the total radius of the cap sputtered on the sphere. Scanning electron microscopy images of \textbf{b}, a $\rm SiO_2-Pd$ and \textbf{b}, $\rm SiO_2-Au$ Janus particles.  The $\rm SiO_2-Au$ is artificially coloured. The scale bar depicts 1.1 $\rm \mu m$}
\label{fig:figS1}
\end{figure}
        
\begin{figure}[H]
\centering
\includegraphics[width=0.8\columnwidth]{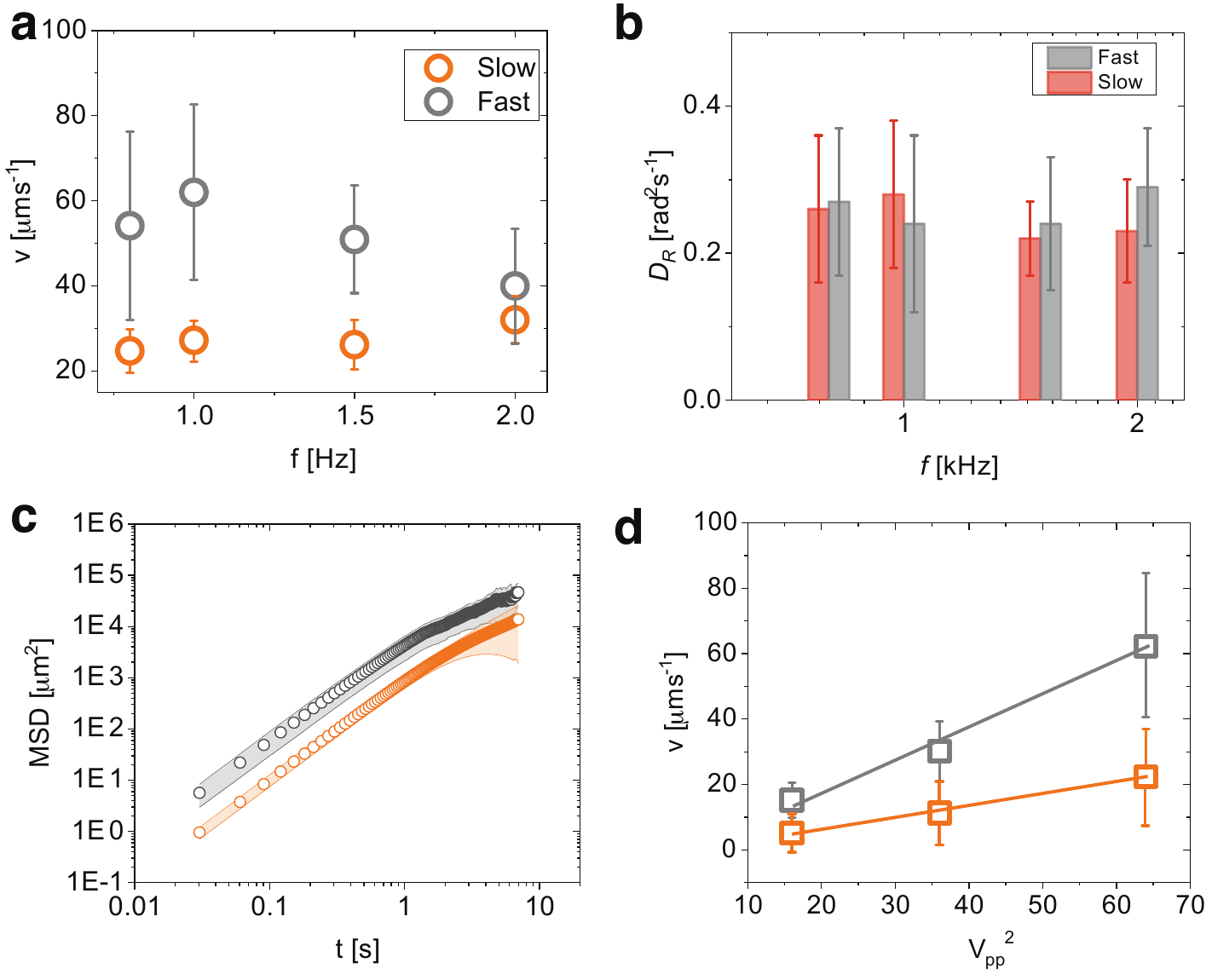}
\caption{\textbf{a}, Velocity as a function of frequency (kHz) at a fixed voltage (8 $\rm V_{pp}$) for the fast $\rm SiO_2-Pd$ (grey) and slow $\rm SiO_2-Au$ (orange) Janus particles, obtained from individual fitting of the short-time mean-squared displacements for more than 1000 trajectories for each species in diluted conditions A\% $\rm \approx 0.01$. The error bar depicts the standard deviation within a 95\% confidence interval. \textbf{b} Rotational diffusivity $\rm D_R$ as a function of frequency for fast and slow particles, obtained from the long-time fitting of the mean-squared displacement in log-log, as $\rm D_R^{-1}= \tau_R$. \textbf{c} Mean-squared displacement at 1 kHz and 8 $\rm V_{pp}$for fast (grey) and slow (orange) Janus particles. The error bars are the standard deviation calculated from independent MSDs. \textbf{c}, Linear dependence of the velocity for fast (grey) and slow (orange) Janus particles as a function of the applied electric field squared at 1 kHz.}
\label{fig:figS2}
\end{figure}

\begin{figure}[H]
\centering
\includegraphics[width=0.8\columnwidth]{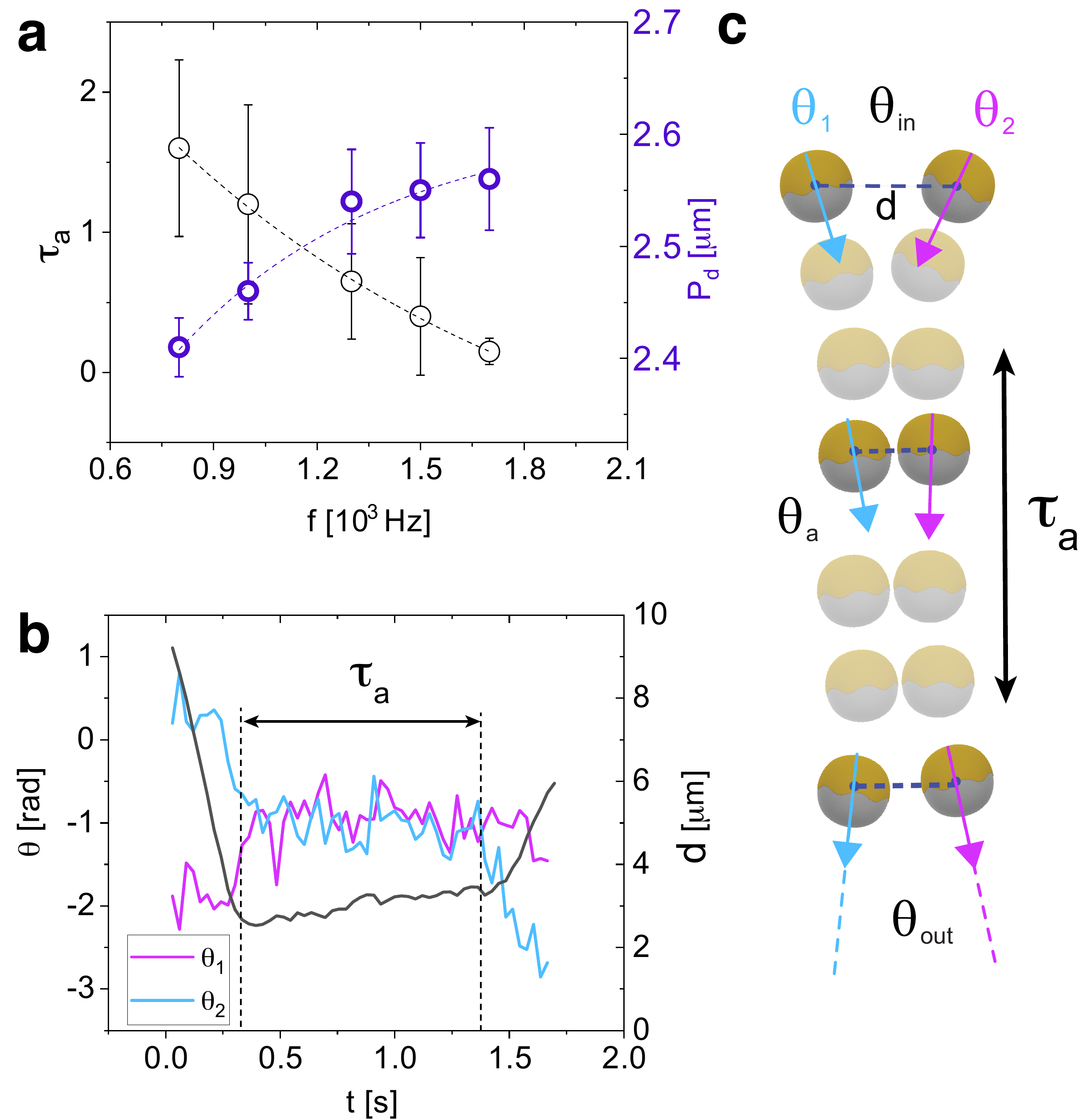}
\caption{\textbf{a}, Alignment time $\rm \tau_{a}$ and inter-particle distance $\rm d$, for individual alignments of two particles as a function of the applied frequency. The data contains both fast and slow Janus particles for more than 50 alignment events. The error bars indicate the standard deviation from the data. The dashed line is a guide to the eye. \textbf{b}, Relative Janus particles orientation $\rm \theta$ and inter-particle distance $\rm d$ as a function of time for a pair of particles undergoing an alignment event at 1 kHz and 8 $\rm V_{pp}$. The alignment time is defined as $\tau_a$,\textbf{c} corresponding scheme of an alignment event between two particles.}
\label{fig:figS3}
\end{figure}

\begin{figure}[H]
\centering
\includegraphics[width=1\columnwidth]{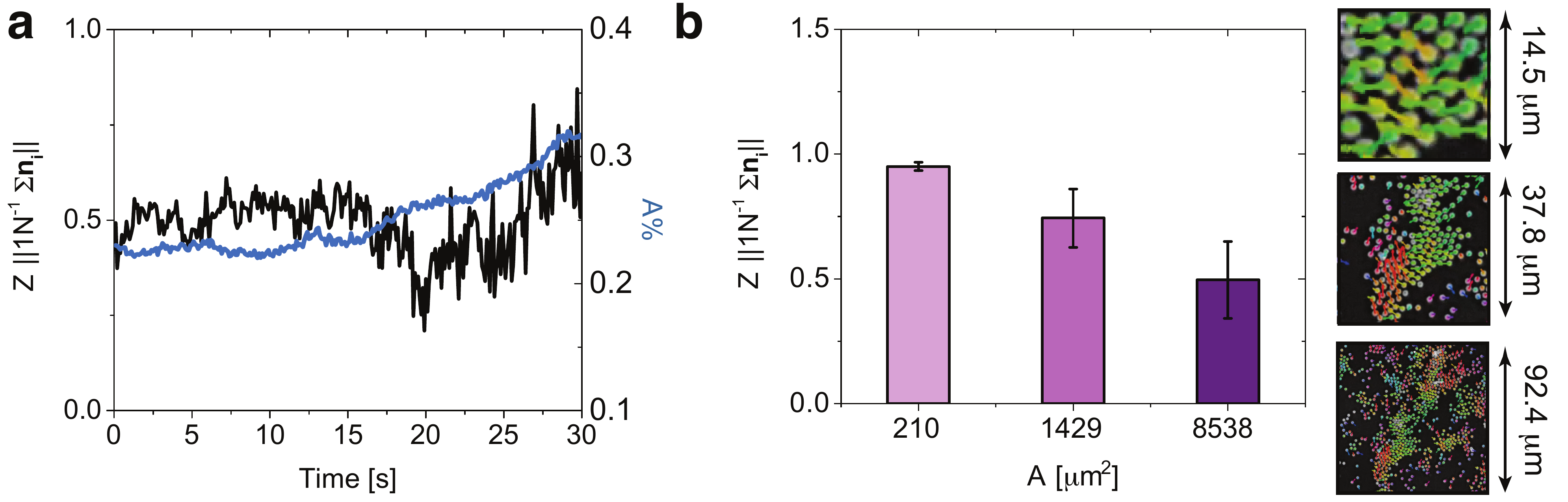}
\caption{ \textbf{a}, Global order parameter  Z and area fraction A\% as a function of time for Movie S4. \textbf{b} Examples of global order Z varying the region of interest (ROI) area, with the corresponding snapshot from experiments at 1 kHz, 8 $\rm V_{pp}$ and A\% $\approx$ 0.2. The highest polar order (Z$\rm \approx$1) is measured at ROI of $\approx$ 200 $\rm \mu m^2$, while as the ROI increases the Z decreases, hinting at local polar order of the system.}
\label{fig:figS4}
\end{figure}

\begin{figure}[H]
\centering
\includegraphics[width=1\columnwidth]{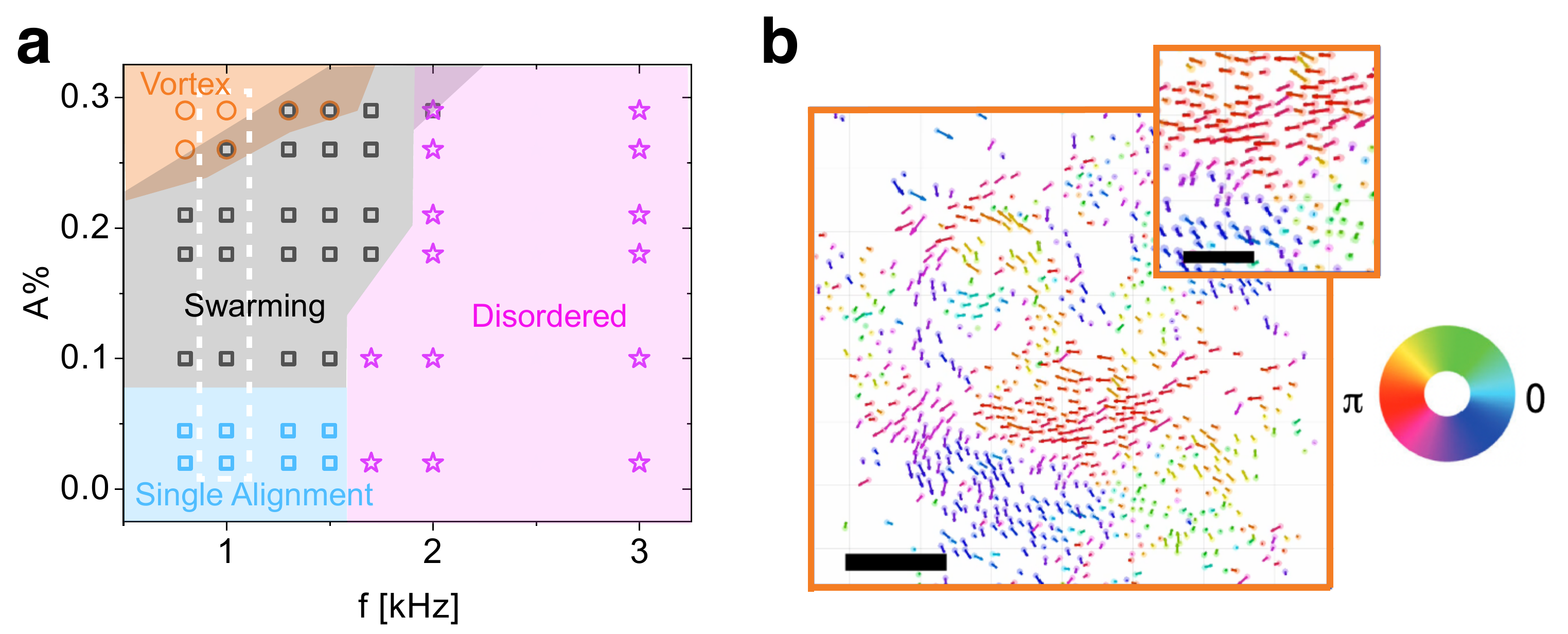}
\caption{\textbf{a}, Experimental dynamical state diagram varying the frequency (0.8 to 3 kHz) and the area fraction A\% (from 0.01 to 0.35).\textbf{b}, Experimental screenshot of vortices observed above 0.25 A\%. Scale bars indicate 100 $\rm \mu m$, and 20 $\rm \mu m$ for the inset. The colours indicate the orientation of the velocity vector.}
\label{fig:figS5}
\end{figure}

\begin{figure}[H]
\centering
\includegraphics[width=1\columnwidth]{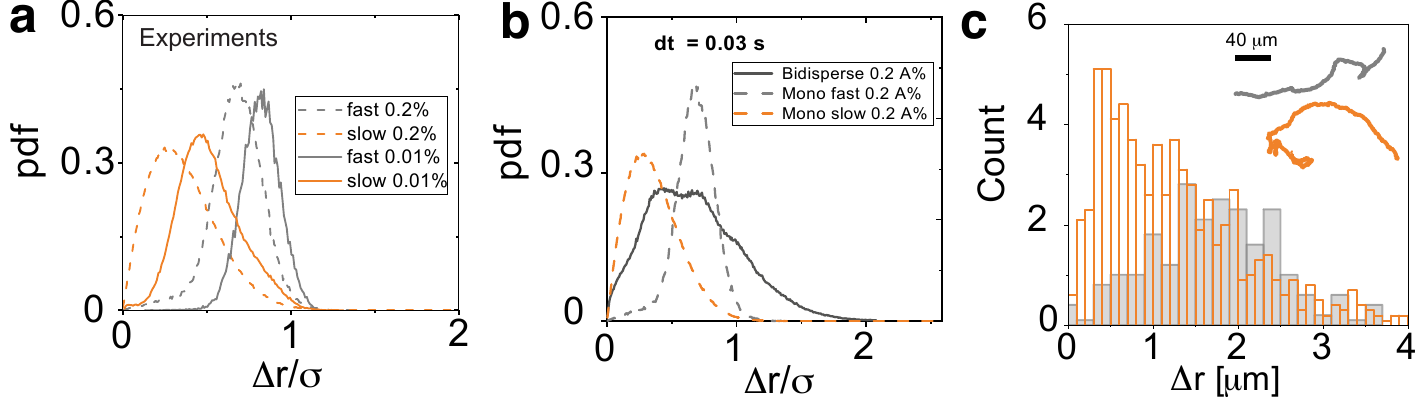}
\caption{ \textbf{a} Experimental probability distribution of displacements for monodisperse samples of fast (grey) and slow (orange) Janus particles at diluted (0.01 A\%, solid line) and dense (0.2 A\%, dashed line) regimes. An effective decrease in the particle displacements due to crowding is observed as expected from the dilute to the dense regimes.\textbf{b}, Comparison of the distribution of displacements between the monodisperse (dashed lines) and bidisperse samples (solid line) containing at 0.2 A\%.  The pdfs are calculated for more than 1000 trajectories, calculated for a dt = 0.3s \textbf{c}, Typical single particle histograms of displacements for slow (orange) and fast (grey) particles in the polar binary mixture phase, with their associated trajectories. }
\label{fig:figS6}
\end{figure}

\section{Fabrication of Janus particles}

Given that the Janus sphere main component is the same $\rm SiO_2$ sphere, we have taken into account the density of each metal sputter onto the $\rm SiO_2$ to match the masses and avoid heavy-bottom effects that might affect the dynamics of the particles.

For this, we calculate the mass associated with a thin cap around the $\rm SiO_2$ sphere, taking into account both the densities for Pt ($\rm \rho_{Pd} = 12 gcm^{-3}$) and Au ($\rm \rho_{Au} = 19.3 gcm^{-3}$). In particular, we first calculate the volume of the cap as

\begin{equation}
    V_{cap} = \frac{2}{3}\pi R^3 - \frac{2}{3}\pi r^3 
\end{equation}

where R is the total radius of the sphere including the cap, and r is the internal radios of the sphere accounting the thickness of the cap. We then calculate the mass of caps of different thicknesses using the different densities of Au and Pt, aiming for the same mass using $\rm m_{cap} = V_{cap} * \rho$.

\section{Dynamical characterization of bidisperse dense mixtures}

We developed a dynamical classification tool to identify fast and slow particles in the bidisperse samples. In the dilute limit, the dynamical distinction remains fairly straightforward, clearly distinguishing fast and slow particles by their instantaneous velocities, which are fitted to a normal distribution. As expected, no interesting phenomena are observed apart from a few alignment events when two particles meet. 

Nonetheless, in the dense phase (see a configuration screenshot in main Fig.3a), the dynamics are highly influenced by the formation of the dynamical polar clusters, where the instantaneous velocity distinction criteria fail to identify fast and slow particles. 
Thus, the classification of fast and slow particles in a cluster of experimental data is based on the analysis of the displacements of each full trajectory. To distinguish fast and slow particles in dense clusters, we analyze the full displacement trajectories of individual particles and compute the total displacements both inside and outside clusters. We construct distributions of these displacements and classify particles as fast or slow based on the shape and fitted parameters of these distributions. In the free state—when particles are not part of a cluster—fast particles exhibit higher total displacements than slow ones. In the clustered state, fast particles show a moderate reduction in displacement due to confinement, while slow particles display increased displacement as a result of being pushed by the fast cluster.
